\documentclass[twocolumn,showpacs,preprintnumbers,amsmath,amssymb,floatfix,pra]{revtex4}
\usepackage{graphicx}
\usepackage{dcolumn}
\usepackage{bm}
\usepackage{amsmath}

\newcommand{\fif}{$^{15}$ND$_3$}
\newcommand{\JK}{$\left|J,K\right>=\left|1,1\right>$}

\begin{document}
\preprint{submitted to PRA}
\title{A versatile electrostatic trap}
\author{Jacqueline van Veldhoven$^{1,2}$}
\author{Hendrick L. Bethlem$^{1,3}$}
\author{Melanie Schnell$^{1}$}
\author{Gerard Meijer$^{1}$}
\affiliation{ ~$^{1}$Fritz-Haber-Institut der Max-Planck-Gesellschaft, Faradayweg 4-6, D-14195 Berlin, Germany\\
~$^{2}$FOM-Institute for Plasmaphysics Rijnhuizen, P.O. Box 1207, NL-3430 BE Nieuwegein, The Netherlands\\
~$^{3}$Laser Centre Vrije Universiteit Amsterdam, de Boelelaan 1081, NL-1081 HV
Amsterdam, The Netherlands}

\date{\today}

\begin{abstract}
A four electrode electrostatic trap geometry is demonstrated that can be used
to combine a dipole, quadrupole and hexapole field. A cold packet of \fif{}
molecules is confined in both a purely quadrupolar and hexapolar trapping field
and additionally, a dipole field is added to a hexapole field to create either
a double-well or a donut-shaped trapping field. The profile of the \fif{} packet
in each of these four trapping potentials is measured, and the dependence of the
well-separation and barrier height of the double-well and donut potential on the
hexapole and dipole term are discussed.
\end{abstract}

 \maketitle
\section{Introduction}
The creation of cold molecules offers unique possibilities for studying cold
collisions
\cite{Burnett:Nat416:225,Balakrishnan:JCP113:621,Balakrishnan:CPL341:652-ohne_url,Bohn:PRA63:052714},
dipole-dipole interaction
\cite{Santos:PRL85:1791,Goral:PRL88:170406,Stuhler:PRL95:150406}, or for doing
high-resolution spectroscopy
\cite{Veldhoven:EPJD31:337-ohne_url,Hudson:arXiv:physics0601054-ohne_url}. In
each of these fields, traps are either necessary, or they enhance the potential
of the method considerably. Electrostatic traps in particular are highly
versatile. They are rather deep (on the order of 1~K) and by changing electrode
geometries, many different trapping potentials can be formed
\cite{Wing:PRL45:631,Bethlem:Nature406:491-ohne_url,Crompvoets:Nat411:174,Meerakker:PRL94:023004-ohne_url,Shafer-Ray:PRA67:045401,Xu:thesis:2001}.
Here we demonstrate an electrode geometry that is used to form different
trapping fields by changing the voltage applied to the electrodes. In the trap
a predominantly dipolar, quadrupolar or hexapolar field can be created. When
implementing the latter two, a pure quadrupole or a pure hexapole trapping
potential is generated. Combining a dipole field with a hexapole field gives
rise to two additional trapping potentials, a double-well or a ring-shaped
potential, depending on the sign of the dipole term. As voltages can be
switched on relatively short timescales, rapid changing between these trapping
fields is feasible as well.

A double-well trap in particular has proven to be fruitful in atomic physics
experiments. For instance, double-well potentials have been used in
interference experiments to study coherence properties of BEC's
\cite{Andrews:Science275:637} and to visualize phase singularities associated
with vortices in a Na BEC \cite{Inouye:PRL87:080402}. Furthermore, two weakly
coupled BEC's in a double-well potential have been shown to form a Josephson
junction \cite{Albiez:PRL95:010402}, and collision studies have been performed in a
double-well interferometer, enabling the determination of the $s$- and $d$-wave
scattering amplitudes of $^{87}\textrm{Rb}$ \cite{Buggle:PRL93:173202}.

In electrostatic traps the formation of a double-well potential with variable
barrier height and well separation is relatively straightforward. Along with
the possibility of rapid switching to other types of trapping fields, this
creates favorable conditions for the study of cold collisions. Molecules
trapped in either of the two wells of a double-well trap will gain potential
energy when switching to a single-well trap, resulting in two packets of
molecules that are accelerated towards each other. As they reach the center of
the trap, they will collide with a velocity that is dependent on the shape of
both the single- and double-well potential. As these shapes can be readily
varied, the velocity of the molecules is tunable. Combining this feature with
an inherently low velocity spread of the trapped molecules makes rapid single-
to double-well conversion an excellent starting point for studying collisions
at low temperatures.

\section{The various trapping configurations}
\begin{figure}
\includegraphics[width=0.7\linewidth,bb= 0 0 200 182]{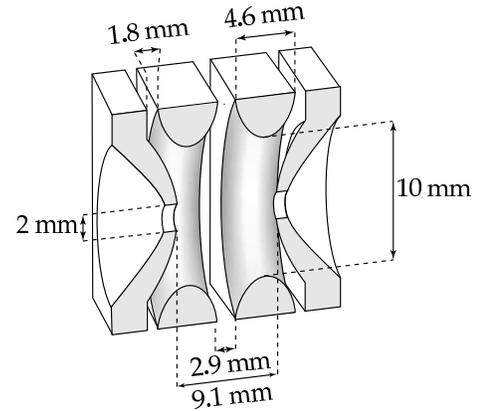}
\caption{Scheme of the trap. The trap is cylindrically symmetric and
consists of two parallel rings in the middle and two endcaps at the sides.}
\label{afmetingen}
\end{figure}
In a cylindrically symmetric geometry, the electric potential $\Phi(\rho,z)$
can be expressed as
\cite{Bergeman:PRA35:1535,Shafer-Ray:PRA67:045401,Xu:thesis:2001}:\footnotesize
\begin{equation}
\displaystyle\Phi(\rho,z) = \Phi_{0} + \Phi_{1}\frac{z}{z_0} +
\Phi_{2}\frac{\left(z^{2} - \rho^{2}/2\right)}{z_0^2} +
\Phi_{3}\frac{\left(z^{3} - 3\rho^{2}z/2\right)}{z_0^3} \dots,\label{Phi}
\end{equation}\normalsize
with $z_0$ a characteristic length scale. Here, the first term is a constant
voltage, the second term a constant electric field, the third a quadrupolar
electric field and the fourth a hexapolar electric field. A molecule with a
quadratic Stark shift will experience a harmonic force in a quadrupolar
electric field, whereas a hexapolar field is needed to form a harmonic trap for
a molecule with a linear Stark shift. In our experiments we use \fif{}
molecules in the low-field seeking hyperfine levels of the \JK{} inversion
doublet of the vibrational and electronic ground state. These molecules
experience a quadratic Stark shift for small electric field strengths and a
linear Stark shift for larger electric field strengths. The dependence of the
Stark shift on the electric field strength $E$ for \fif{} in these levels is
given by \cite{Townes:MicrowaveSpec}:\small
\begin{equation}
W_{Stark}(E)=\sqrt{\left(\frac{W_{inv}}{2}\right)^2+(\frac{1}{2}\mu
E)^2}-\left(\frac{W_{inv}}{2}\right),
\end{equation}\normalsize
with $W_{inv}$ the inversion splitting and $\mu$ the dipole moment. For \fif{}
the inversion splitting is 1430.3~MHz \cite{Veldhoven:EPJD31:337-ohne_url} and
the dipole moment is taken to be the same as for $^{14}$ND$_3$, i.e. 1.48~D
\cite{Bhattacharjee:JMolSpec138:38}. At even larger field strengths, the effect
of mixing with higher rotational levels becomes more important and higher order
terms have to be taken into account.

Using the four electrode geometry shown in Fig.~\ref{afmetingen} it
is possible to create a field with dominant dipolar, quadrupolar or hexapolar
contributions, by applying the appropriate voltages. The geometry of the trap
is the same as was used previously to create an ac electric trap
\cite{Veldhoven:PRL94:083001}. It consists of two rings with an inner diameter
of 10~mm and two endcaps. The two rings have a width of 4.6~mm and are
separated by 2.9~mm, whereas the two endcaps are separated by $2z_0=9.1$~mm.
They each have a hole with a diameter of 2.0~mm, the first to allow molecules
to enter and the second to extract the ions that are produced in the detection
scheme.
\begin{figure}
\includegraphics[width=\linewidth,bb= -10 0 279 438,clip]{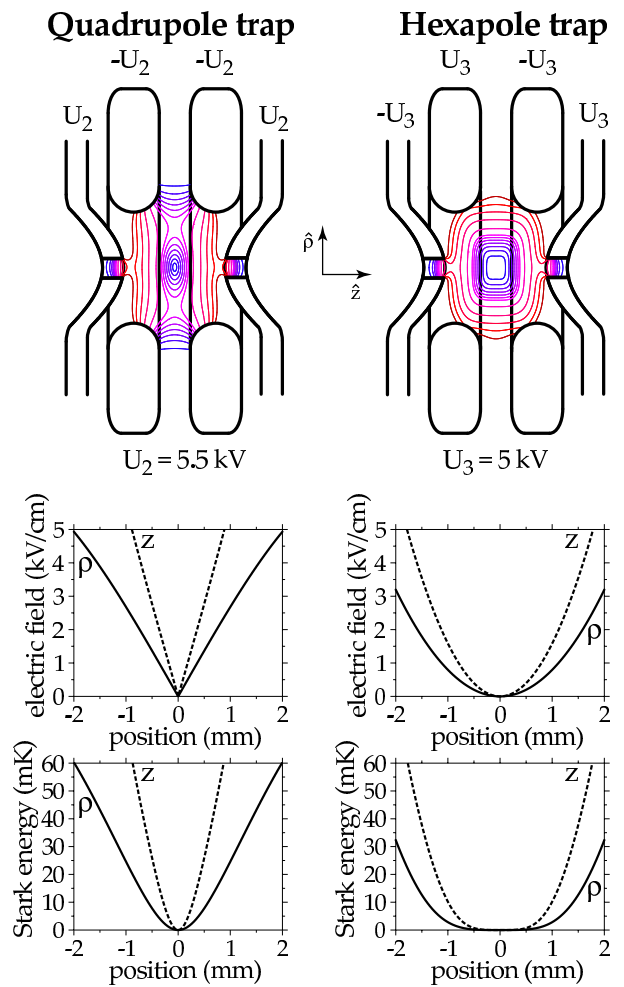}
\caption{Two cross-sections of the trap. On the left-hand side, the voltages
that result in a quadrupole trap are shown, along with lines of constant
electric field for electric field strengths of every kV/cm from 1 to 7~kV/cm
and every 5~kV/cm from 10 to 25~kV/cm (shown in color online, with increasing
field strength going from blue to red). On the right-hand side, the same is
shown for a hexapole trap. Below the two cross-sections, the electric field is
shown along with the Stark energy of a \fif{} molecule as a function of
position in the $\rho$- and z-direction (solid line and dotted line,
respectively).} \label{quadhex}
\end{figure}

In Fig.~\ref{quadhex} the voltages needed for two different trapping fields are
shown. When a voltage $-U_2$ is applied to the two ring electrodes and a
voltage $U_2$ is applied to both endcaps, such as shown on the left-hand side
of the figure, a field is formed with a dominant quadrupole term. Using a
commercially available finite element program \cite{Simion6:1995} the potential
of this configuration is simulated. Below the trap cross-section the electric
field strength as a function of position is shown for both the $\rho$-direction
(solid line) and the z-direction (dotted line). Fitting the corresponding
potential to equation~\ref{Phi}, we find that $\Phi_0\approx -0.43~U_2$,
$\Phi_2\approx 1.05~U_2$, $\Phi_4 \approx 0.35~U_2$, and $\Phi_6\approx
-0.04~U_2$ and that the uneven terms are zero due to symmetry. Beneath the
electric field strength in Fig.~\ref{quadhex} the Stark energy in mK
($T=W_{Stark}/k$, with $k$ the Boltzmann constant) of a \fif{} molecule in this
field is shown. The electric field in the quadrupole trap increases linearly
with distance, but, because of the nonlinear dependence of the Stark energy on
the electric field, the potential is harmonic near the center of the trap.
Further away from the center the potential is linear. The resulting trap is
tight (the trapping frequency is $\omega_z/2\pi$ = 1.7~kHz at the center of the
trap) and has a depth of 0.09~K for \fif{}.

A hexapole trap results when the voltages $U_3$ and $-U_3$ are applied
alternately to the four electrodes of the trap as shown on the right-hand side
of Fig.~\ref{quadhex}. A fit of the potential of this field to
equation~\ref{Phi} yields $\Phi_1=0$, $\Phi_3\approx ~U_3$ and $\Phi_5\approx
-0.04~U_3$, with the even terms being zero due to symmetry. Again, both the
electric field strength along the $\rho$- and z-direction (solid and dotted
line, respectively) and the Stark energy of a \fif{} molecule in this field are
shown below the cross-sections of the trap. The electric field strength in this
trap is quadratically dependent on the distance from the center. However, due
to the nonlinear Stark effect the potential is essentially flat near the center
of the trap. Away from the center the potential becomes harmonic. This trap is
less tight (the trapping frequency is $\omega_z/2\pi$ = 0 at the center of the
trap and goes to $\omega_z/2\pi$ = 750~Hz further away from the center) and has
a depth of 0.35~K for \fif{}.

\begin{figure}[t]
\includegraphics[width=\linewidth,bb= 0 0 289 438,clip]{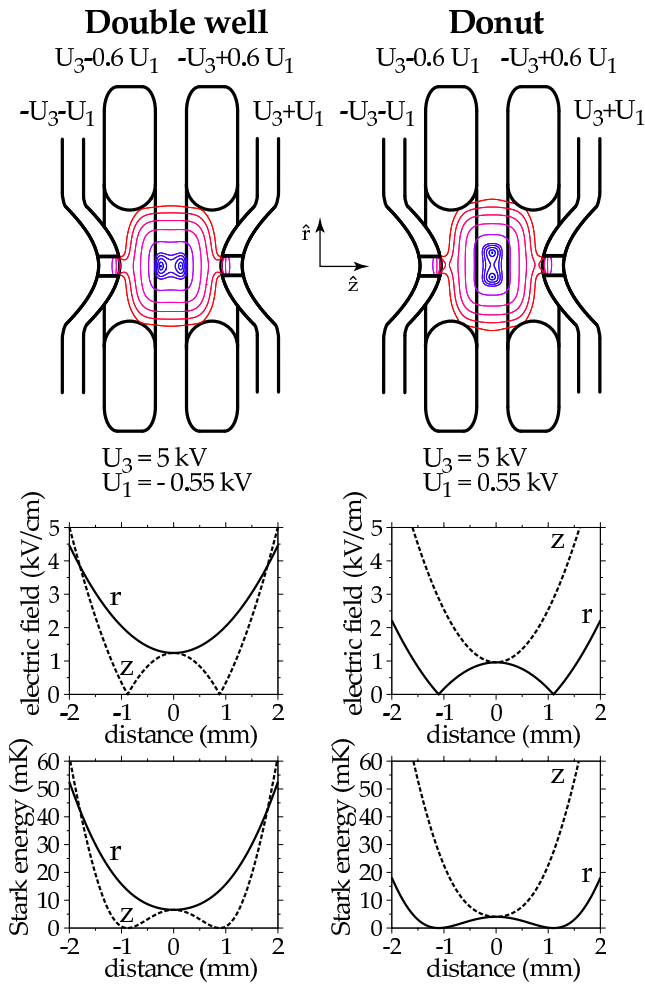}
\caption{Two cross-sections of the trap. On either side the voltages that are
applied when a dipole field is added to a hexapole field are shown. Lines of
constant electric field are shown for electric field strengths of 0.1~kV/cm,
every 0.5~kV/cm from 0.5 to 2~kV/cm, and every 5~kV/cm from 5 to 25~kV/cm
(shown in color online, with increasing field strength going from blue to red).
On the left-hand side (right-hand side) $U_1$ and $U_3$ have the same
(opposite) sign. In each case both the electric field and the Stark energy of a
\fif{} molecule as a function of position are shown in both the $\rho$- and
z-direction (solid line and dotted line, respectively).} \label{dipole}
\end{figure}
\begin{figure}
\includegraphics[width=0.6\linewidth]{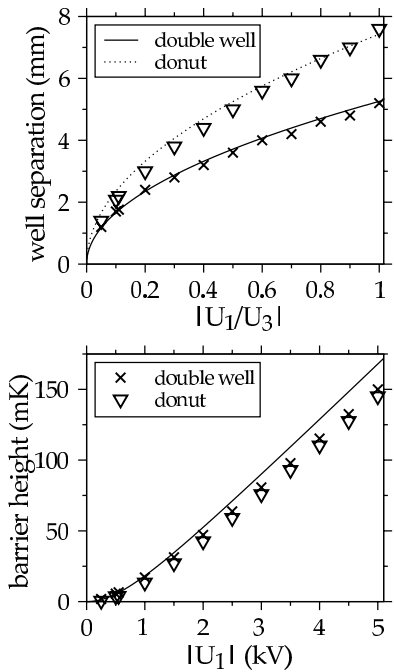}
\caption{Well separation as a function of $|U_1/U_3|$ and barrier height (for
\fif{}) as a function of $|U_1|$ in the ideal double-well (solid line) and
donut trap (dotted line). In both graphs the crosses (double-well trap) and
triangles (donut trap) show the same quantities obtained from electric field
simulations of the actual trap geometry.} \label{wellseparation}
\end{figure}

When applying voltages of -$U_1$, -0.6~$U_1$, 0.6~$U_1$, and $U_1$
consecutively to the four electrodes of the trap, a dipolar field is formed.
Fitting the potential of this field to equation~\ref{Phi} results in
$\Phi_1\approx 0.88~U_1$, $\Phi_3=0$ and $\Phi_5\approx 0.09~U_1$, with the
even terms being zero due to symmetry. Adding a dipole field to a hexapole
field results in either a double-well potential or a donut potential, depending
on the relative sign of the dipole and hexapole term. In Fig.~\ref{dipole} both
configurations are shown, along with the electric field strength and the Stark
energy of a \fif{} molecule in this field along the $\rho$- and z-direction
(solid and dotted line, respectively). Both the distance between the two minima
in the double-well trap and the diameter of the ring-shaped minimum in the
donut trap are dependent on the ratio between the dipole and hexapole term. In
the upper graph of Fig.~\ref{wellseparation} the distance between the two wells
is shown as a function of the ratio $|U_1/U_3|$ in the case of an ideal
hexapole and dipole field (solid line). The dotted line shows the same for the
diameter of the donut-shaped minimum. The crosses and triangles (double-well
trap and donut trap, respectively) indicate the simulated value of this well
separation, using the finite element program with $U_3=5$~kV and different
values of $U_1$.

The height of the barrier between the two wells is shown in the lower graph of
Fig.~\ref{wellseparation}, given in mK for \fif{}. In the case of an ideal
hexapole and dipole field (solid line), this height is the same for the
double-well and donut trap and is solely dependent on the value of $|U_1|$.
Note that the barrier height is identical to the amount of potential energy
gained by a \fif{} molecule in the center of one of the wells when switching to
the single-well configuration. This can be seen by realizing that in the center
of the double-well trap, i.e. at the top of the barrier, the electric field
strength is determined by the dipole term only, as the hexapole field is zero
there. In the center of one of the wells the hexapole field is equal to the
dipole field, although opposite in direction. Therefore, when switching the
dipole field off to produce a single-well trap, molecules originally at the
center of one of the two wells will now have a potential energy equal to the
barrier height. Values for the barrier height were obtained using the finite
element program as well and are shown as crosses (double-well trap) and
triangles (donut trap) in the lower graph of Fig.~\ref{wellseparation}. From
these two graphs it is clear that both the well separation and the barrier
height, and therefore also the collision energy of the molecules upon double-
to single-well conversion, can easily be changed by varying the voltages on the
four electrodes. At the maximum value of $|U_1|$ used in the experiment, $|U_1|
= 0.55$~kV, the barrier height is on the order of 5~mK. The separation between
the wells is 1.8~mm in this case ($U_3=5$~kV), whereas the diameter of the
donut-shaped minimum is 2.2~mm.

\section{Experiment}
\begin{figure}
\includegraphics[width=\linewidth,bb = 0 0 286 143]{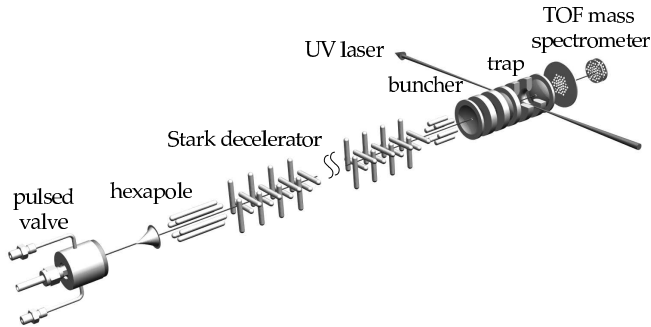}
\caption{Experimental setup. A pulsed supersonic beam of \fif{} molecules
seeded in Xenon exits a cooled valve with a velocity of 280~m/s. Ammonia
molecules in the low-field seeking hyperfine levels of the \JK{} state are
focused into the decelerator by a hexapole, decelerated, and then transversally
and longitudinally focused into the trap by a hexapole and buncher,
respectively. The molecules are detected in a TOF-mass spectrometer.}
\label{experiment}
\end{figure}
In Fig.~\ref{experiment} the experimental setup is shown. In the experiments,
the trap is loaded from a Stark-decelerated molecular beam. A mixture of 5~\%
\fif{} seeded in Xenon is supersonically expanded from a pulsed valve at a
10~Hz rate. Due to cooling of the valve to $-70\,^{\circ}\textrm{C}$, the gas
pulse has a velocity of 280~m/s. In the expansion, about 60~\% of the ammonia
molecules are internally cooled to the ground state of para-ammonia, the \JK{}
inversion doublet in the electronic and vibrational ground state. After passing
through a skimmer, those molecules that reside in the low-field seeking
hyperfine levels of the upper component of the inversion doublet are focused
into the Stark decelerator by a 6~cm long hexapole. Molecules in high-field
seeking levels are attracted by the high electric fields of both the hexapoles
and the decelerator and will be lost from the beam. Molecules that reside in
the remainder of the hyperfine levels of the inversion doublet, that only have
a higher order Stark effect, will hardly be effected by the electric fields.
The decelerator consists of 95 pairs of parallel 3~mm-diameter rods, with a
closest distance of 2~mm between two rods and a closest distance of 2.5~mm
between pairs of rods. Each consecutive pair of rods is rotated over
$90\,^{\circ}$ and within each pair, a positive voltage of 10~kV is applied to
one electrode, whereas the same negative voltage is applied to the other. A
more extensive description of this part of the setup, along with the operation
principle of the decelerator can be found elsewhere
\cite{Veldhoven:EPJD31:337-ohne_url,Bethlem:PRA65:053416}. Using a phase angle
of $57.5^{\circ}$ a subset of low-field seeking molecules with an initial
velocity of around 280~m/s is decelerated to about 15~m/s at the exit of the
decelerator. After 5~mm the molecules are transversally focused into the trap
by a second (12.5~mm long) hexapole. Subsequently, they are longitudinally
focused into the trap using a buncher that is placed another 5~mm downstream of
the hexapole. The buncher has a similar geometry as the trap. Before the
ammonia molecules enter the trap, voltages are applied to the trap such that a
last electric field slope is formed. In gaining Stark energy when entering the
trap, the low-field seeking molecules lose their last kinetic energy and reach
a stand-still in the middle of the trap. At this moment the voltages on the
four electrodes are switched to generate a particular trapping potential,
thereby confining the molecules in either a quadrupole, hexapole, double-well
or donut potential. After a certain trapping time, the trap is turned off and
molecules in the upper component of the inversion doublet are ionized in a
state-selective (2+1)-Resonance Enhanced Multi Photon Ionization (REMPI)
scheme, using a pulsed UV-laser around 317~nm \cite{Ashfold:JCP89:1754}. The
resulting ions are subsequently detected in a time-of-flight (TOF)
mass-spectrometer. Using a 75~cm lens the ionization laser is focused into the
trap passing through the 2.9~mm gap between the two ring electrodes. The laser
focus is estimated to be about 200~$\mu$m in diameter. Applying small bias
voltages of 200~V, 150~V, -150~V and -200~V to the four electrodes of the trap
results in an electric field that accelerates the ions towards the detector.
The ions are detected over a length of 2~mm along the laserbeam, limited by the
size of the hole in the endcap. Using a computer controlled translation stage,
the focus of the laserbeam can be moved along the beam-axis, i.e. along the
z-direction, over about 2.5~mm. In this way, the spatial distribution of the
trapped molecules can be measured.

\section{Results and discussion}
\begin{figure}
\includegraphics[width=0.5\linewidth]{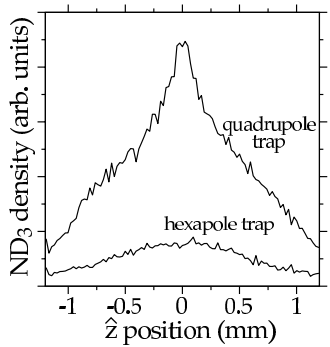}
\caption{Density of trapped \fif{} molecules as a function of position in the
z-direction for both the quadrupole trap (upper trace) and the hexapole trap
(lower trace), after trapping for about 72~ms.} \label{compare}
\end{figure}
In Fig.~\ref{compare}, the density of trapped \fif{} molecules is shown as a
function of position in the z-direction, both for hexapole and for quadrupole
trapping voltages. The measurements are taken after the molecules have been
trapped for about 72~ms. We found that in order to obtain a single-well
profile for the molecules confined in the hexapole trap a small dipole term
had to be added. We attribute this to deviations of the distances between the
electrodes from the ideal geometry. This compensating dipole term is added in
all subsequent measurements. As the quadrupole trap is much tighter than the
hexapole trap (see Fig.~\ref{quadhex}), a signal difference arises between the
two traces. The peak density of molecules in the hexapole trap is about $10^7
\textrm{cm}^{-3}$, whereas it is about ten times higher in the quadrupole trap.
The loading process is identical for both traps and is matched to the hexapole
trap, resulting in a less than perfect match of the decelerated packet to the
quadrupole trap. This gives rise to the peculiar shape of the measured
distribution in the quadrupole trap, which seems to consist of two packets with
different temperatures.

\begin{figure}
\includegraphics[width=\linewidth]{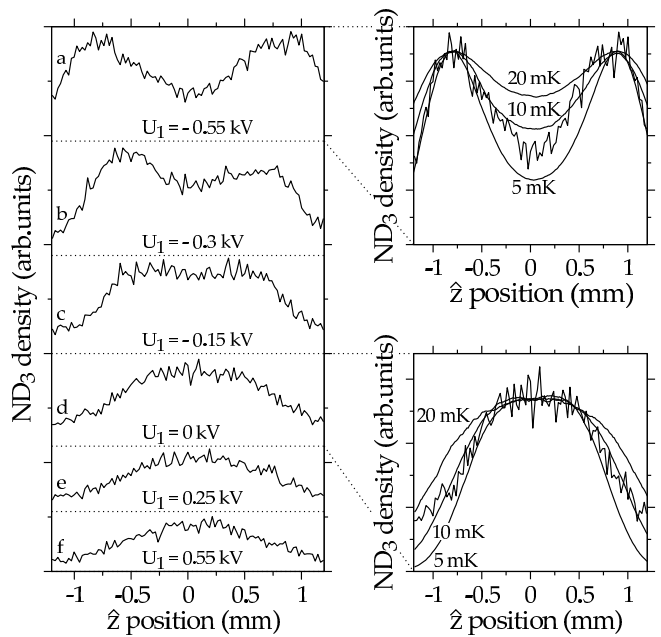}
\caption{Density of \fif{} molecules confined in a hexapole trap as a function
of position in the z-direction. A negative dipole term of -0.55~kV, -0.3~kV and
-0.15 k~V is added to the hexapole term ($U_3 = 5$~kV) in the upper three
traces (a-c), resulting in a double-well potential. Trace (d) shows a packet in
a purely single well. A positive dipole term of 0.25~kV and 0.55~kV is added
for the lower two traces (e-f), thereby creating a donut potential. On the
right-hand side, trace (a) and (d) are shown once more, along with simulations
of the density distribution of the packet of molecules for temperatures of 5,
10 and 20~mK.} \label{kameeldonut}
\end{figure}
In Fig.~\ref{kameeldonut} the density of \fif{} molecules in the hexapole trap
is shown as a function of position in the z-direction. By adding a dipole term
to the hexapole term, the shape of the trap and thereby the profile of the
trapped molecules can be changed significantly. Each trace in
Fig.~\ref{kameeldonut} corresponds to a different value of the added dipole
term. When no dipole is added, the trap consists of only a single well. A
single peak results, as trace (d) shows. Adding a negative dipole term
transforms the single well to a double well, with a distance and barrier
between the two wells that depend on the value of the dipole term. The upper
three traces of Fig.~\ref{kameeldonut} show two peaks that move closer together
as the dipole term is increased from -0.55~kV (trace (a)) to -0.3~kV (trace
(b)) to -0.15~kV (trace (c)). When a positive dipole term is added, the shape
of the trap becomes that of a donut. As only molecules in the relatively small
laser focus near the center of the trap are ionized in the detection scheme,
this transformation to a donut-shaped trap leads to a reduction in signal only,
as is shown by trace (e) and (f) ($U_1$ = 0.25~kV and $U_1$ = 0.55~kV,
respectively).

\begin{figure}
\includegraphics[width=\linewidth]{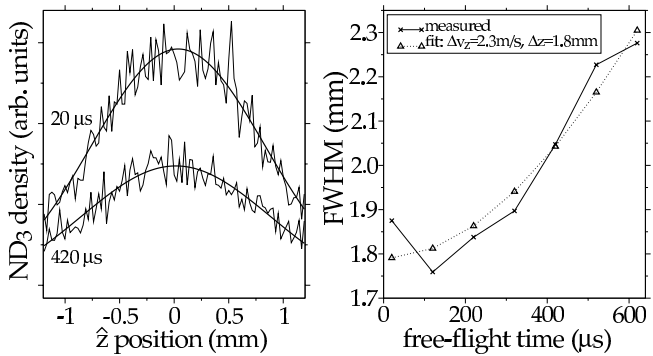}
\caption{Determination of the temperature of a packet of \fif{} molecules
trapped in the hexapole trap. On the left, the profile of a packet is shown
20~$\mu$s and 420~$\mu$s after the trap is turned off (upper and lower trace,
respectively). From a fit to a Gaussian lineshape the FWHM of the distribution
is obtained, as shown as a function of free-flight time (solid line with
crosses) on the right-hand side of the figure.} \label{hextemp}
\end{figure}

At the densities that we have in the trap, no thermalizing collisions occur,
and formally no temperature can be assigned. To nevertheless characterize
the trapped sample in terms of a temperature, we fitted the density profile
to the thermal density distribution \cite{Luiten:PRA53:381}:
\begin{equation}
n(\rho,z)=n_0e^{-W_{Stark}(\rho,z)/kT},\label{density}
\end{equation}
with $n_0$ the density at the trap minimum and $W_{Stark}(\rho,z)$ the
potential energy as a function of position. On the right-hand side of
Fig.~\ref{kameeldonut} the measurements with a dipole term of $U_1 = -0.55$~kV
and $U_1$ = 0~kV are shown once more, along with the thermal density
distribution as obtained from equation~(\ref{density}) for temperatures of 5,
10 and 20~mK. It can be seen that in either case the simulated distribution
with a temperature of 10~mK matches the measurement best. The same simulation
has been performed for the spatial distribution of molecules confined in a
quadrupole trap. Due to the seemingly double profile, it is more difficult to
describe this distribution with a single temperature. A temperature of 75~mK
seems to best describe the broad base, whereas the sharp center peak agrees
best with a temperature of 5~mK. An alternative method to deduce a temperature
is to measure the width of the packet of trapped molecules as a function of
time after the trap has been turned off. Measurements of this type are shown in
Fig.~\ref{hextemp}. On the left-hand side two profiles of \fif{} molecules are
seen, one after 20~$\mu$s and one after 420~$\mu$s after switching off the
hexapole trap (upper and lower curve, respectively). In either case the
molecules have been confined for about 72~ms. Due to the velocity spread of the
molecules the packet spreads out after the trap has been turned off, resulting
in less signal and an increased width. To obtain this width a Gaussian
lineshape is fitted to each profile. The acquired full-width-half-maximum
(FWHM) is shown as a function of free-flight time on the right-hand side of the
figure (solid line with crosses). To deduce a temperature, this curve is fitted
to $\sqrt{(\Delta z)^2+(\Delta v_z\cdot t)^2}$ (dotted line with triangles),
resulting in a velocity spread of 2.3~m/s and a position spread of 1.8~mm (both
FWHM). Using $\Delta
v_z=\sqrt{8\mathrm{ln}2}\cdot\sqrt{kT/m}$~\cite{Reif:1985-2}, this gives a
temperature of about 2.5~mK.

The different temperatures of 10~mK and 2.5~mK obtained by the two different
methods actually correspond to a difference of only a few tenths of a mm in the
width of the initial packet. Such a small difference can be explained by
possible misalignments of the electrodes of the trap, causing uncertainties in
the actual electric field, and therefore in $W_{Stark}(\rho,z)$. Moreover,
extracting the width of the packets from the measurements is quite sensitive to
how the baseline is taken into account in the fit. All together, this analysis
only serves to indicate that the temperature of the trapped sample is in the mK
range.

\begin{figure}
\includegraphics[width=0.5\linewidth]{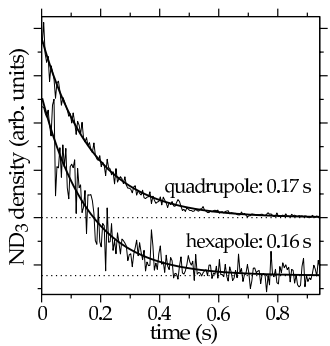}
\caption{Density of \fif{} molecules in the quadrupole (upper trace) and the
hexapole (lower trace) trap as a function of time that the trap is on.}
\label{lifetime}
\end{figure}
In Fig.~\ref{lifetime} the density of ammonia molecules is shown in both the
quadrupole and the hexapole trap as a function of time that the trap is on. The
molecules are seen to leave the trap with a $1/e$ time of 0.16~s for the
hexapole trap and 0.17~s for the quadrupole trap, the same within error bars.
The trapping lifetime is mainly limited by background collisions (background
pressure $4\cdot10^{-8}$~mbar). Optical pumping by black-body radiation cannot
be completely neglected, however. In the room-temperature trap, pumping of the
\fif{} molecules to the vibrationally excited $\nu_2$ level (umbrella mode)
around 750~cm$^{-1}$ is expected to occur on the time-scale of a second.

\section{Conclusions}
We have presented a new electrostatic trapping geometry, that illustrates the
versatility of electric fields in forming tailored trapping potentials. By
applying different voltages to a set of four electrodes, different trapping
fields are created, between which rapid switching on a timescale of 100~ns is
possible. Spatial distributions and lifetimes of a cold packet of \fif{}
molecules loaded into both a hexapole and a quadrupole trap are shown.
Additionally, we studied a hexapole trap that transforms into a
double-well or donut trap when different dipole terms are added. The barrier
height and well separation of either one of these two traps depend on the
strength of the dipolar and hexapolar electric field and are easily changed.

Rapid switching between, for instance, the double-well trap and a pure hexapole
trap offers good prospects for measuring collision cross-sections as a function
of collision energy at low temperatures. The amount of energy gained by
molecules trapped in one of the wells after switching to the single-well
depends solely on the strength of the electric dipole field, which can be
easily varied.

\begin{acknowledgments}  This work is part of the research program of the
`Stichting voor Fundamenteel Onderzoek der Materie (FOM)', which is financially
supported by the Netherlands Organisation for Scientific Research (NWO). H.L.B.
acknowledges financial support from NWO via a VENI-grant. The research of M.S.
was made possible through a Liebig stipend of the `Fonds der Chemischen
Industrie'. We thank Andr\'e J.A. van Roij and Henrik Haak for the mechanical
design and construction of the trap and Boris Sartakov for helpful discussions.
\end{acknowledgments}
\bibliography{string,mp}

\begin{thebibliography}{28}
\expandafter\ifx\csname natexlab\endcsname\relax\def\natexlab#1{#1}\fi
\expandafter\ifx\csname bibnamefont\endcsname\relax
  \def\bibnamefont#1{#1}\fi
\expandafter\ifx\csname bibfnamefont\endcsname\relax
  \def\bibfnamefont#1{#1}\fi
\expandafter\ifx\csname citenamefont\endcsname\relax
  \def\citenamefont#1{#1}\fi
\expandafter\ifx\csname url\endcsname\relax
  \def\url#1{\texttt{#1}}\fi
\expandafter\ifx\csname urlprefix\endcsname\relax\def\urlprefix{URL }\fi
\providecommand{\bibinfo}[2]{#2}
\providecommand{\eprint}[2][]{\url{#2}}

\bibitem[{\citenamefont{Burnett et~al.}(2002)\citenamefont{Burnett, Julienne,
  Lett, Teisinga, and Williams}}]{Burnett:Nat416:225}
\bibinfo{author}{\bibfnamefont{K.}~\bibnamefont{Burnett}},
  \bibinfo{author}{\bibfnamefont{P.~S.} \bibnamefont{Julienne}},
  \bibinfo{author}{\bibfnamefont{P.~D.} \bibnamefont{Lett}},
  \bibinfo{author}{\bibfnamefont{E.}~\bibnamefont{Teisinga}}, \bibnamefont{and}
  \bibinfo{author}{\bibfnamefont{C.~J.} \bibnamefont{Williams}},
  \bibinfo{journal}{Nature} \textbf{\bibinfo{volume}{416}},
  \bibinfo{pages}{225} (\bibinfo{year}{2002}).

\bibitem[{\citenamefont{Balakrishnan and
  Dalgarno}(2000)}]{Balakrishnan:JCP113:621}
\bibinfo{author}{\bibfnamefont{N.}~\bibnamefont{Balakrishnan}}
  \bibnamefont{and} \bibinfo{author}{\bibfnamefont{A.}~\bibnamefont{Dalgarno}},
  \bibinfo{journal}{J. Chem. Phys.} \textbf{\bibinfo{volume}{113}},
  \bibinfo{pages}{621} (\bibinfo{year}{2000}).

\bibitem[{\citenamefont{Balakrishnan and
  Dalgarno}(2001)}]{Balakrishnan:CPL341:652-ohne_url}
\bibinfo{author}{\bibfnamefont{N.}~\bibnamefont{Balakrishnan}}
  \bibnamefont{and} \bibinfo{author}{\bibfnamefont{A.}~\bibnamefont{Dalgarno}},
  \bibinfo{journal}{Chem. Phys. Lett.} \textbf{\bibinfo{volume}{341}},
  \bibinfo{pages}{652} (\bibinfo{year}{2001}).

\bibitem[{\citenamefont{Bohn}(2001)}]{Bohn:PRA63:052714}
\bibinfo{author}{\bibfnamefont{J.~L.} \bibnamefont{Bohn}},
  \bibinfo{journal}{Phys. Rev. A} \textbf{\bibinfo{volume}{63}},
  \bibinfo{pages}{052714} (\bibinfo{year}{2001}).

\bibitem[{\citenamefont{Santos et~al.}(2000)\citenamefont{Santos, Shlyapnikov,
  Zoller, and Lewenstein}}]{Santos:PRL85:1791}
\bibinfo{author}{\bibfnamefont{L.}~\bibnamefont{Santos}},
  \bibinfo{author}{\bibfnamefont{G.~V.} \bibnamefont{Shlyapnikov}},
  \bibinfo{author}{\bibfnamefont{P.}~\bibnamefont{Zoller}}, \bibnamefont{and}
  \bibinfo{author}{\bibfnamefont{M.}~\bibnamefont{Lewenstein}},
  \bibinfo{journal}{Phys. Rev. Lett.} \textbf{\bibinfo{volume}{85}},
  \bibinfo{pages}{1791} (\bibinfo{year}{2000}).

\bibitem[{\citenamefont{G\'oral et~al.}(2002)\citenamefont{G\'oral, Santos, and
  Lewenstein}}]{Goral:PRL88:170406}
\bibinfo{author}{\bibfnamefont{K.}~\bibnamefont{G\'oral}},
  \bibinfo{author}{\bibfnamefont{L.}~\bibnamefont{Santos}}, \bibnamefont{and}
  \bibinfo{author}{\bibfnamefont{M.}~\bibnamefont{Lewenstein}},
  \bibinfo{journal}{Phys. Rev. Lett.} \textbf{\bibinfo{volume}{88}},
  \bibinfo{pages}{170406} (\bibinfo{year}{2002}).

\bibitem[{\citenamefont{Stuhler et~al.}(2005)\citenamefont{Stuhler, Griesmaier,
  Koch, Fattori, Pfau, Giovanazzi, Pedri, and Santos}}]{Stuhler:PRL95:150406}
\bibinfo{author}{\bibfnamefont{J.}~\bibnamefont{Stuhler}},
  \bibinfo{author}{\bibfnamefont{A.}~\bibnamefont{Griesmaier}},
  \bibinfo{author}{\bibfnamefont{T.}~\bibnamefont{Koch}},
  \bibinfo{author}{\bibfnamefont{M.}~\bibnamefont{Fattori}},
  \bibinfo{author}{\bibfnamefont{T.}~\bibnamefont{Pfau}},
  \bibinfo{author}{\bibfnamefont{S.}~\bibnamefont{Giovanazzi}},
  \bibinfo{author}{\bibfnamefont{P.}~\bibnamefont{Pedri}}, \bibnamefont{and}
  \bibinfo{author}{\bibfnamefont{L.}~\bibnamefont{Santos}},
  \bibinfo{journal}{Phys. Rev. Lett.} \textbf{\bibinfo{volume}{95}},
  \bibinfo{pages}{150406} (\bibinfo{year}{2005}).

\bibitem[{\citenamefont{van Veldhoven et~al.}(2004)\citenamefont{van Veldhoven,
  K\"upper, Bethlem, Sartakov, van Roij, and
  Meijer}}]{Veldhoven:EPJD31:337-ohne_url}
\bibinfo{author}{\bibfnamefont{J.}~\bibnamefont{van Veldhoven}},
  \bibinfo{author}{\bibfnamefont{J.}~\bibnamefont{K\"upper}},
  \bibinfo{author}{\bibfnamefont{H.~L.} \bibnamefont{Bethlem}},
  \bibinfo{author}{\bibfnamefont{B.}~\bibnamefont{Sartakov}},
  \bibinfo{author}{\bibfnamefont{A.~J.} \bibnamefont{van Roij}},
  \bibnamefont{and} \bibinfo{author}{\bibfnamefont{G.}~\bibnamefont{Meijer}},
  \bibinfo{journal}{Eur. Phys. J. D} \textbf{\bibinfo{volume}{31}},
  \bibinfo{pages}{337} (\bibinfo{year}{2004}).

\bibitem[{\citenamefont{Hudson et~al.}(2006)\citenamefont{Hudson, Lewandowski,
  Sawyer, and Ye}}]{Hudson:arXiv:physics0601054-ohne_url}
\bibinfo{author}{\bibfnamefont{E.~R.} \bibnamefont{Hudson}},
  \bibinfo{author}{\bibfnamefont{H.~J.} \bibnamefont{Lewandowski}},
  \bibinfo{author}{\bibfnamefont{B.~C.} \bibnamefont{Sawyer}},
  \bibnamefont{and} \bibinfo{author}{\bibfnamefont{J.}~\bibnamefont{Ye}},
  \bibinfo{journal}{arXiv} p. \bibinfo{pages}{physics/0601054}
  (\bibinfo{year}{2006}).

\bibitem[{\citenamefont{Wing}(1980)}]{Wing:PRL45:631}
\bibinfo{author}{\bibfnamefont{W.~H.} \bibnamefont{Wing}},
  \bibinfo{journal}{Phys. Rev. Lett.} \textbf{\bibinfo{volume}{45}},
  \bibinfo{pages}{631} (\bibinfo{year}{1980}).

\bibitem[{\citenamefont{Bethlem et~al.}(2000)\citenamefont{Bethlem, Berden,
  Crompvoets, Jongma, van Roij, and Meijer}}]{Bethlem:Nature406:491-ohne_url}
\bibinfo{author}{\bibfnamefont{H.~L.} \bibnamefont{Bethlem}},
  \bibinfo{author}{\bibfnamefont{G.}~\bibnamefont{Berden}},
  \bibinfo{author}{\bibfnamefont{F.~M.~H.} \bibnamefont{Crompvoets}},
  \bibinfo{author}{\bibfnamefont{R.~T.} \bibnamefont{Jongma}},
  \bibinfo{author}{\bibfnamefont{A.~J.~A.} \bibnamefont{van Roij}},
  \bibnamefont{and} \bibinfo{author}{\bibfnamefont{G.}~\bibnamefont{Meijer}},
  \bibinfo{journal}{Nature} \textbf{\bibinfo{volume}{406}},
  \bibinfo{pages}{491} (\bibinfo{year}{2000}).

\bibitem[{\citenamefont{Crompvoets et~al.}(2001)\citenamefont{Crompvoets,
  Bethlem, Jongma, and Meijer}}]{Crompvoets:Nat411:174}
\bibinfo{author}{\bibfnamefont{F.~M.~H.} \bibnamefont{Crompvoets}},
  \bibinfo{author}{\bibfnamefont{H.~L.} \bibnamefont{Bethlem}},
  \bibinfo{author}{\bibfnamefont{R.~T.} \bibnamefont{Jongma}},
  \bibnamefont{and} \bibinfo{author}{\bibfnamefont{G.}~\bibnamefont{Meijer}},
  \bibinfo{journal}{Nature} \textbf{\bibinfo{volume}{411}},
  \bibinfo{pages}{174} (\bibinfo{year}{2001}).

\bibitem[{\citenamefont{van~de Meerakker et~al.}(2005)\citenamefont{van~de
  Meerakker, Smeets, Vanhaecke, Jongma, and
  Meijer}}]{Meerakker:PRL94:023004-ohne_url}
\bibinfo{author}{\bibfnamefont{S.~Y.~T.} \bibnamefont{van~de Meerakker}},
  \bibinfo{author}{\bibfnamefont{P.~H.~M.} \bibnamefont{Smeets}},
  \bibinfo{author}{\bibfnamefont{N.}~\bibnamefont{Vanhaecke}},
  \bibinfo{author}{\bibfnamefont{R.~T.} \bibnamefont{Jongma}},
  \bibnamefont{and} \bibinfo{author}{\bibfnamefont{G.}~\bibnamefont{Meijer}},
  \bibinfo{journal}{Phys. Rev. Lett.} \textbf{\bibinfo{volume}{94}},
  \bibinfo{pages}{023004} (\bibinfo{year}{2005}).

\bibitem[{\citenamefont{Shafer-Ray et~al.}(2003)\citenamefont{Shafer-Ray,
  Milton, Furneaux, Abraham, and Kalbfleisch}}]{Shafer-Ray:PRA67:045401}
\bibinfo{author}{\bibfnamefont{N.~E.} \bibnamefont{Shafer-Ray}},
  \bibinfo{author}{\bibfnamefont{K.~A.} \bibnamefont{Milton}},
  \bibinfo{author}{\bibfnamefont{B.~R.} \bibnamefont{Furneaux}},
  \bibinfo{author}{\bibfnamefont{E.~R.~I.} \bibnamefont{Abraham}},
  \bibnamefont{and} \bibinfo{author}{\bibfnamefont{G.~R.}
  \bibnamefont{Kalbfleisch}}, \bibinfo{journal}{Phys. Rev. A}
  \textbf{\bibinfo{volume}{67}}, \bibinfo{pages}{045401}
  (\bibinfo{year}{2003}).

\bibitem[{\citenamefont{Xu}(2001)}]{Xu:thesis:2001}
\bibinfo{author}{\bibfnamefont{G.}~\bibnamefont{Xu}}, Ph.D. thesis,
  \bibinfo{school}{The University of Texas at Austin} (\bibinfo{year}{2001}).

\bibitem[{\citenamefont{Andrews et~al.}(1997)\citenamefont{Andrews, Townsend,
  Miesner, Durfee, Kurn, and Ketterle}}]{Andrews:Science275:637}
\bibinfo{author}{\bibfnamefont{M.~R.} \bibnamefont{Andrews}},
  \bibinfo{author}{\bibfnamefont{C.~G.} \bibnamefont{Townsend}},
  \bibinfo{author}{\bibfnamefont{H.-J.} \bibnamefont{Miesner}},
  \bibinfo{author}{\bibfnamefont{D.~S.} \bibnamefont{Durfee}},
  \bibinfo{author}{\bibfnamefont{D.~M.} \bibnamefont{Kurn}}, \bibnamefont{and}
  \bibinfo{author}{\bibfnamefont{W.}~\bibnamefont{Ketterle}},
  \bibinfo{journal}{Science} \textbf{\bibinfo{volume}{275}},
  \bibinfo{pages}{637} (\bibinfo{year}{1997}).

\bibitem[{\citenamefont{Inouye et~al.}(2001)\citenamefont{Inouye, Gupta,
  Rosenband, Chikkatur, G\"orlitz, Gustavson, Leanhardt, Pritchard, and
  Ketterle}}]{Inouye:PRL87:080402}
\bibinfo{author}{\bibfnamefont{S.}~\bibnamefont{Inouye}},
  \bibinfo{author}{\bibfnamefont{S.}~\bibnamefont{Gupta}},
  \bibinfo{author}{\bibfnamefont{T.}~\bibnamefont{Rosenband}},
  \bibinfo{author}{\bibfnamefont{A.~P.} \bibnamefont{Chikkatur}},
  \bibinfo{author}{\bibfnamefont{A.}~\bibnamefont{G\"orlitz}},
  \bibinfo{author}{\bibfnamefont{T.~L.} \bibnamefont{Gustavson}},
  \bibinfo{author}{\bibfnamefont{A.~E.} \bibnamefont{Leanhardt}},
  \bibinfo{author}{\bibfnamefont{D.~E.} \bibnamefont{Pritchard}},
  \bibnamefont{and} \bibinfo{author}{\bibfnamefont{W.}~\bibnamefont{Ketterle}},
  \bibinfo{journal}{Phys. Rev. Lett.} \textbf{\bibinfo{volume}{87}},
  \bibinfo{pages}{080402} (\bibinfo{year}{2001}).

\bibitem[{\citenamefont{Albiez et~al.}(2005)\citenamefont{Albiez, Gati,
  F\"olling, Hunsmann, Cristiani, and Oberthaler}}]{Albiez:PRL95:010402}
\bibinfo{author}{\bibfnamefont{M.}~\bibnamefont{Albiez}},
  \bibinfo{author}{\bibfnamefont{R.}~\bibnamefont{Gati}},
  \bibinfo{author}{\bibfnamefont{J.}~\bibnamefont{F\"olling}},
  \bibinfo{author}{\bibfnamefont{S.}~\bibnamefont{Hunsmann}},
  \bibinfo{author}{\bibfnamefont{M.}~\bibnamefont{Cristiani}},
  \bibnamefont{and} \bibinfo{author}{\bibfnamefont{M.~K.}
  \bibnamefont{Oberthaler}}, \bibinfo{journal}{Phys. Rev. Lett.}
  \textbf{\bibinfo{volume}{95}}, \bibinfo{pages}{010402}
  (\bibinfo{year}{2005}).

\bibitem[{\citenamefont{Buggle et~al.}(2004)\citenamefont{Buggle, L\'eonard,
  von Klitzing, and Walraven}}]{Buggle:PRL93:173202}
\bibinfo{author}{\bibfnamefont{C.}~\bibnamefont{Buggle}},
  \bibinfo{author}{\bibfnamefont{J.}~\bibnamefont{L\'eonard}},
  \bibinfo{author}{\bibfnamefont{W.}~\bibnamefont{von Klitzing}},
  \bibnamefont{and} \bibinfo{author}{\bibfnamefont{J.~T.~M.}
  \bibnamefont{Walraven}}, \bibinfo{journal}{Phys. Rev. Lett.}
  \textbf{\bibinfo{volume}{93}}, \bibinfo{pages}{173202}
  (\bibinfo{year}{2004}).

\bibitem[{\citenamefont{Bergeman et~al.}(1987)\citenamefont{Bergeman, Erez, and
  Metcalf}}]{Bergeman:PRA35:1535}
\bibinfo{author}{\bibfnamefont{T.}~\bibnamefont{Bergeman}},
  \bibinfo{author}{\bibfnamefont{G.}~\bibnamefont{Erez}}, \bibnamefont{and}
  \bibinfo{author}{\bibfnamefont{H.~J.} \bibnamefont{Metcalf}},
  \bibinfo{journal}{Phys. Rev. A} \textbf{\bibinfo{volume}{35}},
  \bibinfo{pages}{1535} (\bibinfo{year}{1987}).

\bibitem[{\citenamefont{Townes and Schawlow}(1975)}]{Townes:MicrowaveSpec}
\bibinfo{author}{\bibfnamefont{C.~H.} \bibnamefont{Townes}} \bibnamefont{and}
  \bibinfo{author}{\bibfnamefont{A.~L.} \bibnamefont{Schawlow}},
  \emph{\bibinfo{title}{Microwave Spectroscopy}} (\bibinfo{publisher}{Dover
  Publications}, \bibinfo{address}{New York}, \bibinfo{year}{1975}).

\bibitem[{\citenamefont{Bhattacharjee et~al.}(1989)\citenamefont{Bhattacharjee,
  Johnston, Sudhakaran, and Sarker}}]{Bhattacharjee:JMolSpec138:38}
\bibinfo{author}{\bibfnamefont{R.~L.} \bibnamefont{Bhattacharjee}},
  \bibinfo{author}{\bibfnamefont{L.~H.} \bibnamefont{Johnston}},
  \bibinfo{author}{\bibfnamefont{G.~R.} \bibnamefont{Sudhakaran}},
  \bibnamefont{and} \bibinfo{author}{\bibfnamefont{J.~C.}
  \bibnamefont{Sarker}}, \bibinfo{journal}{J. Mol. Spec.}
  \textbf{\bibinfo{volume}{138}}, \bibinfo{pages}{38} (\bibinfo{year}{1989}).

\bibitem[{\citenamefont{van Veldhoven et~al.}(2005)\citenamefont{van Veldhoven,
  Bethlem, and Meijer}}]{Veldhoven:PRL94:083001}
\bibinfo{author}{\bibfnamefont{J.}~\bibnamefont{van Veldhoven}},
  \bibinfo{author}{\bibfnamefont{H.~L.} \bibnamefont{Bethlem}},
  \bibnamefont{and} \bibinfo{author}{\bibfnamefont{G.}~\bibnamefont{Meijer}},
  \bibinfo{journal}{Phys. Rev. Lett.} \textbf{\bibinfo{volume}{94}},
  \bibinfo{pages}{083001} (\bibinfo{year}{2005}).

\bibitem[{\citenamefont{Dahl}(1995)}]{Simion6:1995}
\bibinfo{author}{\bibfnamefont{D.~A.} \bibnamefont{Dahl}},
  \emph{\bibinfo{title}{{\it Simion 3D Version 6.0}}}
  (\bibinfo{publisher}{Idaho National Engineering Laboratory},
  \bibinfo{address}{Idaho Falls (USA)}, \bibinfo{year}{1995}).

\bibitem[{\citenamefont{Bethlem et~al.}(2002)\citenamefont{Bethlem, Crompvoets,
  Jongma, van~de Meerakker, and Meijer}}]{Bethlem:PRA65:053416}
\bibinfo{author}{\bibfnamefont{H.~L.} \bibnamefont{Bethlem}},
  \bibinfo{author}{\bibfnamefont{F.~M.~H.} \bibnamefont{Crompvoets}},
  \bibinfo{author}{\bibfnamefont{R.~T.} \bibnamefont{Jongma}},
  \bibinfo{author}{\bibfnamefont{S.~Y.~T.} \bibnamefont{van~de Meerakker}},
  \bibnamefont{and} \bibinfo{author}{\bibfnamefont{G.}~\bibnamefont{Meijer}},
  \bibinfo{journal}{Phys. Rev. A} \textbf{\bibinfo{volume}{65}},
  \bibinfo{pages}{053416} (\bibinfo{year}{2002}).

\bibitem[{\citenamefont{Ashfold et~al.}(1988)\citenamefont{Ashfold, Dixon,
  Little, Stickland, and Western}}]{Ashfold:JCP89:1754}
\bibinfo{author}{\bibfnamefont{M.~N.~R.} \bibnamefont{Ashfold}},
  \bibinfo{author}{\bibfnamefont{R.~N.} \bibnamefont{Dixon}},
  \bibinfo{author}{\bibfnamefont{N.}~\bibnamefont{Little}},
  \bibinfo{author}{\bibfnamefont{R.~J.} \bibnamefont{Stickland}},
  \bibnamefont{and} \bibinfo{author}{\bibfnamefont{C.~M.}
  \bibnamefont{Western}}, \bibinfo{journal}{J. Chem. Phys.}
  \textbf{\bibinfo{volume}{89}}, \bibinfo{pages}{1754} (\bibinfo{year}{1988}).

\bibitem[{\citenamefont{Luiten et~al.}(1996)\citenamefont{Luiten, Reynolds, and
  Walraven}}]{Luiten:PRA53:381}
\bibinfo{author}{\bibfnamefont{O.~J.} \bibnamefont{Luiten}},
  \bibinfo{author}{\bibfnamefont{M.~W.} \bibnamefont{Reynolds}},
  \bibnamefont{and} \bibinfo{author}{\bibfnamefont{J.~T.~M.}
  \bibnamefont{Walraven}}, \bibinfo{journal}{Phys. Rev. A}
  \textbf{\bibinfo{volume}{53}}, \bibinfo{pages}{381} (\bibinfo{year}{1996}).

\bibitem[{\citenamefont{Reif}(1985)}]{Reif:1985-2}
\bibinfo{author}{\bibfnamefont{F.}~\bibnamefont{Reif}},
  \emph{\bibinfo{title}{Fundamentals of statistical and thermal physics}}
  (\bibinfo{publisher}{McGraw-Hill}, \bibinfo{address}{Singapore},
  \bibinfo{year}{1985}).

\end{thebibliography}
\end{document}